# Bootstrapping Learned Cost Models with Synthetic SQL Queries (Extended Abstract)

Michael Nidd, Christoph Miksovic, Thomas Gschwind,
Francesco Fusco, Andrea Giovannini, Ioana Giurgiu
{mni,cmi,thg,ffu,agv,igi}@zurich.ibm.com
IBM Research Europe

## ABSTRACT

Having access to realistic workloads for a given database instance is extremely important to enable stress and vulnerability testing, as well as to optimize for cost and performance. Recent advances in learned cost models have shown that when enough diverse SQL queries are available, one can effectively and efficiently predict the cost of running a given query against a specific database engine. In this paper, we describe our experience in exploiting modern synthetic data generation techniques, inspired by the generative AI and LLM community, to create high-quality datasets enabling the effective training of such learned cost models. Initial results show that we can improve a learned cost model's predictive accuracy by training it with 45% fewer queries than when using competitive generation approaches.

## 1 INTRODUCTION

Composable data management systems [24] decouple queries from the underlying engines, enabling scenarios in which accurately predicting the best engine to run a query can bring significant performance benefits and cost savings. Learned cost models (LCM) [2, 3, 12, 16, 22, 23, 34] have been recently proposed as a natural solution to this complex optimization problem. When trained on realistic workloads, LCMs are able to accurately match queries with the database engine providing the best performance, resulting in substantial cost optimizations [27].

The availability of such high-quality datasets represents the main obstacle for the effective training and deployment of LCMs. Queries provided by commonly available benchmarks [1], such as TPC-H, TPC-DS or IMDB, can be used to compile training datasets to train LCMs. Still, the queries used in those benchmarks are not representative of every workload and, in addition, are limited in size and diversity. In fact, it is well known that database benchmarks do not accurately reflect real-world queries, and are generally too simple [4, 25, 30]. To alleviate this issue, some approaches [19, 33] employ mechanical synthetic query generators, which typically result in a high volume of queries similar to those in the seed benchmark. Others [32] use historical queries to predict future performance. Such an approach assumes that queries repeat frequently, which is not always the case. Thus, to ensure enough diversity, a system would need to be bootstrapped, i.e., record a substantial volume of queries over a long period of time to train the LCM successfully. During the bootstrapping phase, the system would need to run with suboptimal performance, leading to user dissatisfaction, or at higher costs due to over-provisioning.

In this paper, we propose practical methods to alleviate the bootstrapping problem using modern synthetic data generation techniques based on Large Language Models (LLMs). Recent advances of LLMs and, in particular, of code models enabled the creation of Text2SQL systems offering a natural language interface to query database systems [13, 26]. Using natural language is attractive as it allows non-expert users to easily extract insights from the data stored in a SQL database.

In our scenario, we use the capabilities of LLMs to generate datasets of queries that correspond to a given database schema. Our goal is to generate a set of queries that is not only *diverse*, but also *large* and *representative* enough to train an LCM effectively for a given schema. By measuring the execution time of each query over several database engines, we create representative training sets for such an LCM.

To evaluate our approach, we perform a detailed analysis of the diversity of the generated data and, additionally, we measure the predictive power of the LCM proposed in [27] when trained using data created with our methodology and compare it against the same model trained with mechanically generated queries. The results highlight that our approach enables the training of models that are more accurate and yield an improved query routing, while using *45%* fewer samples.

## 2 BACKGROUND

Recent advances in LLMs have shown that synthetic data generation (SDG) plays a fundamental role in improving LLMs for a broad set of tasks, including coding. The majority of recent LLMs are trained with synthetic data (e.g., LLama3 [7]) and major AI vendors are providing SDG toolkits to create synthetic data to fine-tune models for vertical domains (e.g., IBM Watsonx.ai[9] and Meta LLama[28]).

Synthetic data generation has been fueling the recent advancements of text-to-SQL systems [5, 20, 21, 35]. In contrast with SQL generation used for testing [11, 15], where the objective is to test databases for vulnerabilities and execution bugs using the fewest SQL queries, we are interested in creating high-quality datasets that are sufficiently large and diverse to be used for tuning, benchmarking and, eventually, to train learned cost models.

We leverage synthetic data generation to build a dataset of realistic and diverse SQL queries targeting a specific database instance and schema, eliminating the need for manual annotation. Unlike synthetic data generation used for fine-tuning LLMs, which requires generating data across diverse databases to prevent overfitting, our approach focuses on generating data specific to a particular database schema.

## 3 SCALABLE SYNTHETIC DATA GENERATION

To generate the data for a given database schema, we build on our experience in delivering Synthetic Data Generation (SDG) pipelines for NL-to-SQL tasks in a commercial system and the corresponding open-source version called DiGiT[8]. DiGiT allows us to generate a set of diverse and high-quality SQL queries to train ML-based cost models predicting what would be the cost of running an incoming query over a specific execution engine.

Our SDG pipeline comprises three complementary data generation components: LLM-based, template-based, and SQL-component-based generators, providing a diverse range of generated queries. The quality of the generated data is controlled using *validators*, which are responsible to score and filter generated SQL queries according to multiple criteria including SQL syntax correctness, similarity to existing queries (e.g., to quantify the diversity of the generated data), semantic consistency between natural language and SQL queries. Validators are instrumental in enabling a self-instruct approach [31]. In a self-instruct setup, data is generated in an iterative way using an external LLM to enlarge the dataset with synthetically generated, but validated data. Self-instructing enables the creation of more diverse training datasets, breaking away from the structure of the initial seed examples.

To use the framework, a user must provide a database schema, along with optional metadata, seed examples that showcase typical utterances and queries to be generated, as well as optional prompt rules. For template and component-based data generation, access to a database allows for including sampled tables and columns as well as concrete values from the database (e.g., filter conditions of the SQL query). Furthermore, the framework calculates several metrics for the generated samples, including SQL complexity, SQL functions, and coverage of tables and columns. These are meant to assess the quality and diversity of the generated data.

The SDG framework offers multiple approaches to generate diverse data samples. One common method, applicable to our scenario, involves processing metadata from external data sources, such as schema descriptions in Database Description Language (DDL), data catalogs, or other relevant sources. The framework leverages this schema information, which can be enriched with annotations, data value categories, and ranges, to produce natural language utterances and corresponding SQL code. Additionally, it can generate synthetic data by combining schema information with query log data, which provides valuable insights into user behavior by capturing executed queries across various applications. In this scenario, natural language utterances are generated for the given queries, resulting in a triplet comprising schema, query, and utterance. In some cases, such triples may be readily available or created by domain experts, referred to as ground truth or seed examples. This data can be used as few-shot examples, alongside the database schema, to generate data samples that mimic the style of the provided exemplars.

Several parameters can be adjusted to control the SQL code generation process: (1) utilize different database sub-schemas to focus on particular tables and columns, (2) generate and prioritize few-shot examples based on identified coverage gaps, (3) improve LLM prompts by refining textual generation rules and adaptively adjusting constraints based on validation feedback, and (4) variation of LLM generation parameters such as temperature or top-p to increase creativity or penalizing repeated tokens to improve output quality by reducing redundancy.

To assess the quality of synthetically generated SQL queries, several statistical measures are applied after they have passed validation checks for query syntactic correctness and semantic equivalence. The analysis involves checking for key SQL clauses, logical operators, and aggregate functions, and providing a distribution of these features across the generated queries. Additionally, the evaluation tracks the frequency of references to each table and column in the underlying database schema, offering insights into schema coverage.

## 4 STEERING SDG TOWARDS DIVERSITY

Our synthetic query data generator is based on the pipeline in Figure 1. The pipeline consists of the following steps:

***Preprocessing.*** This step extracts the raw schema from the database, including table and view information and column metadata such as column types, primary keys, or foreign keys. However, not all metadata is declared, and some metadata, such as enumerations (e.g., day names) or version numbers (e.g., "3.0.1") cannot be derived from the schema directly. In the future, we will explore using LLMs to identify this metadata as well as its benefits in the data generation.

***Create Subschema.*** Next, the query generator creates targeted subschemas (connectable subsets of tables from the full database schema). From our work on DiGiT (§3), we found that SQL generation performs best when prompts include a targeted subschema. This context improves the precision of the generated queries and helps steer generation toward semantically meaningful joins and filters. Accordingly, this step constructs a large number of subschema groups connected by foreign keys, either explicitly declared in the schema or inferred during preprocessing.

***DataBuilder.*** The DataBuilder loops through the created subschemas and creates a prompt for each instance. The prompt style is "few-shot", providing the schema in the form of a CREATE statement for each table, and then listing a few example queries before requesting that a fresh query be generated. This corresponds to the LLM-method based on schema metadata and seed examples described in Section 3.

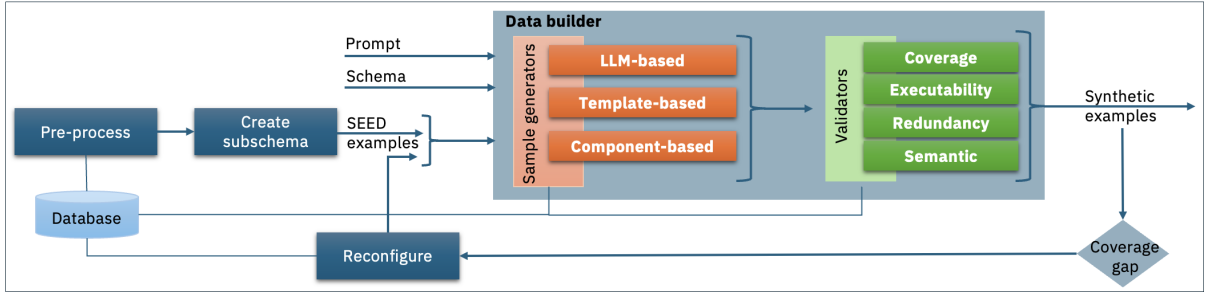

Figure 1: SDG pipeline with input and output (SQL queries) artifacts.

The example queries are built mechanically (i.e. algorithmically) from the subschemas. They use simple configurable pseudo-random selection to create valid queries that include SQL constructs such as *group by* operations or filters. This is flexible enough to guide selections toward areas with poor coverage, but does not generate the variety of query construction that we obtain in combination with the LLM. If coverage reports already exist from previous iterations, this will direct the generation to fill any coverage gaps that exist by including only a targeted subset of columns in each table definition, and by tuning the example queries towards specific operations. Finally, we specify LLM parameters [18] to generate multiple different responses for each request.

The returned queries are then filtered and deduplicated by validators based on correct syntax and relevance. While syntax filtering is simple, relevance has a great deal of freedom. For example, if preprocessing identified that a column contains only two possible values, then this will be reflected both in the examples in the prompt, and also used in the relevance filtering. Similarly, arithmetic operations on columns representing labels such as version numbers are filtered out.

*Coverage.* To augment the standard metrics provided by the SDG framework, we use Apache Calcite [10] to parse SQL statements and combine the resulting syntax tree with the known schema. This allows us to identify how often tables and columns are referenced, the frequency of operations, the structure and conditions of joins, and the use of sub-selects. By counting these occurrences, this coverage step ensures that data is generated up front with broad coverage and high internal consistency. If gaps are identified, we reconfigure the pipeline and bias the generation of additional queries. This approach contrasts with a previous study on quantifying data coverage [14], where we used a vector of metrics to assess the quality and distribution of human-generated data. Since generating new data was costly and impractical, we instead identified subsets with high internal consistency, trained separate models for each, and had to discard the rest of the data.

## 5 RESULTS

Using the TPC-H schema [6], the current pipeline implementation generates 187 possible groupings of tables (without repetition). We refer to each of these groupings as subschemas, and pass them into a `granite-3.3-8b-instruct` model [29], as described in Section 4. Multiple generations are produced for each prompt to increase variability. Moving forward, we will also be performing quantitative performance comparisons between generative models.

The base prompt follows the template outlined in Figure 2. In our experiments, we have also tested intentional bias towards *group by* and *order by* operators by adding a prompt constraint, such as "Whenever possible, please use a group by clause. Use operators for more complex groups." before listing the seed examples. In experiments that included seed examples, we also weighted the example generation to include the desired clause with 90% probability (see §A.1 Fig.5).

We used the mechanically-generated examples as the baseline to assess how much the generative model improves on the characteristics of the queries and eventually on query execution time predictions. To this end, we applied two distinct settings: 1) one not including examples in the prompt (0-shot), and 2) including 3 few-shot examples (3-shot) as well as the mentioned prompt constraints.

*Coverage of Structural Complexity Categories.* To analyze the structural complexity of the generated queries, we group them into complexity buckets based on SQL keyword occurrences. These include joins, standard clauses, logical operators, and functions.

```
These tables have been created:
<List of CREATE statements>
Write an interesting and complicated SQL query
that uses all of these tables:
<List of table names>
These are some examples:
<Enumerated list of SELECT statements>
```

Figure 2: Generation prompt outline

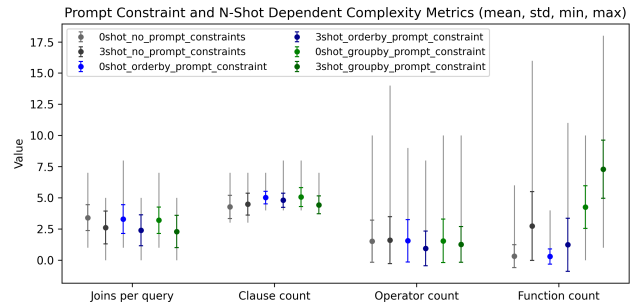

Figure 3: Query structure by generation method.

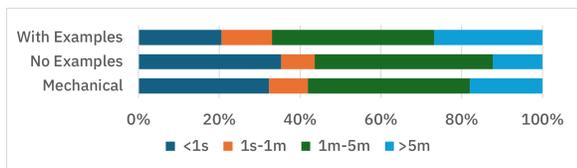

Figure 4: Runtime characteristics of generated queries.

Each of these buckets represents a different facet of query complexity. Here we compare six "prompt settings", defined by combinations of few-shot examples (0-shot vs. 3-shot) and prompt constraints (*no* vs. *order by* vs. *group by* bias). Figure 3 visualizes the distribution for each complexity bucket across prompt settings, showing mean, standard deviation, minimum, and maximum values. This allows us to assess how the different prompting strategies influence query complexity across multiple dimensions. One clear trend is that the number of joins per query tends to decrease when 3-shot examples are used. This likely reflects an inductive bias toward simpler patterns introduced by the examples themselves. Clause and operator counts remain relatively stable across settings. However, a slight reduction can be observed when prompt constraints like *order by* or *group by* are combined with 3-shot examples. While these changes are subtle but consistent, they may suggest that prompt constraints encourage more concise SQL formulations. Function usage, in contrast, is strongly affected by both the presence of prompt constraints and the number of seed examples. Across all prompt settings, 3-shot prompts lead to a strong increase in the number of functions used, suggesting that exposure to examples encourages richer functional expressions. In addition, prompts that request *group by* already induce more function usage even in the 0-shot setting. This is likely due to the inherent association between grouping and aggregation. These are promising results, demonstrating variety in the resulting SQL statements that will become even more useful when we progress to queries with more complex structures like sub-selects, for which variety is harder to achieve using the mechanical algorithm.

***Query Execution Time & Cardinality.*** Figure 4 demonstrates the range of actual execution times for the generated queries. In this experiment, the TPC-H database was restricted to a maximum 40K rows per table, and queries were run with a timeout of 10 minutes. The results are grouped into four buckets, $< 1 second$, $1 second - 1 minute$, $1 - 5 minutes$, and $> 5 minutes$. Prompts requesting bias towards *group by* and *order by* are run both without and with seed examples (0-shot, 3-shot). The prompts with seed examples have a more even distribution across all buckets. This uniformity of run time can be expected to provide the most representative group of test queries for measuring overall performance (see §A.2 Fig.6).

***Bootstrapping Learned Cost Models.*** Our ultimate goal in using LLMs to generate synthetic queries is to actually improve the accuracy of LCMs for the downstream task of optimizing query routing towards the most beneficial engine in a lakehouse. To that end, in [27] we proposed a multi-predictor head GNN-based LCM that predicts query execution times for a variety of engines supported in the typical lakehouse. We refer the reader to [27] for details on the LCM's architecture and training setup.

|  | $q_{median}$ | $q_{mean}$ | $q_{p95}$ |
|---|---|---|---|
| $\text{Err}_{\text{GNN+Mech}}$ | 1.20 | 1.41 | 2.38 |
| $\text{Err}_{\text{GNN+SDG}}$ | 1.18 | 1.34 | 2.37 |

Table 1: LCM accuracy in predicting query execution times.

We evaluate the LCM on Spark-SQL and PrestoDB engines, each provisioned with 1 and 4 worker nodes, respectively, by training it on 4000 mechanically generated queries based on the TPC-H benchmark and testing it on 1000 queries. The technique employed to generate the queries has been proposed in [12]. The model's accuracy, $\text{Err}_{\text{GNN+Mech}}$, shown in Table 1, was measured by using the Q-error metric, averaged over all considered engine types and provisionings. The metric $q_{median}$ for a set of queries $Q$ and a set of engines $\mathcal{E}$ is computed as:

$$q_{median} = \frac{1}{|\mathcal{E}|} \sum_{i \in 1...|\mathcal{E}|} \text{median}\left(\left\{\max\left(\frac{pred_q}{true_q}, \frac{true_q}{pred_q}\right) : q \in Q\right\}\right)$$

A median Q-error of 1.20 means that, on average 50% of the estimates deviate by no more than 20% from the true, measured execution time. In the case of a perfect predictor, Q-error = 1.

Given that the mechanical data generator used to train the LCM includes random joins, predicates, and aggregations without concerns for diversity across functions and operators, our intuition is that training the LCM on SQL queries generated by this SDG pipeline will improve the LCM's accuracy. Indeed, we observe that even when trained on only 2200 queries (45% fewer queries than the baseline), $\text{Err}_{\text{GNN+SDG}}$ decreases slightly for $q_{median}$ and $q_{p95}$ and more significantly for $q_{mean}$ (i.e., a 0.07 drop). More detailed results for each of the considered engines and their respective provisionings are included in §A.3 Table 2. Furthermore, we evaluate the actual query routing across the considered engines based on the predictions of the LCM trained when trained with the 2200 queries generated with SDG. In comparison to the case where 4000 mechanically generated queries are used for training, the LCM+SDG provide a reduction from 165 min to 150 min for the 1000 test queries (10% improvement). These initial results show that using an SDG to train an LCM improves query execution time prediction and routing even with significantly fewer generated queries.

## 6 CONCLUSIONS AND OUTLOOK

We have demonstrated an innovative LLM-based approach for generating a diverse and balanced collection of SQL queries that are suitable for training an LCM, whose ultimate goal is to aid in the optimal routing of workloads across various engine flavors and provisionings. Even with 45% fewer training samples, the LCM is able to reduce its prediction error when compared to the case where mechanically generated queries are used for the training step, which results in an improved actual query routing.

These initial results encourage us to further improve the SDG pipeline by optimizing the prompts and the LLM through a reinforcement learning loop informed by the LCM's learned representations, in addition to the validator metrics that reflect query diversity and operator/function coverage. We are also experimenting with other generator models and exploring the use of DSPy [17] to programatically improve the prompt.

# A APPENDIX

## A.1 Coverage of Structural Complexity Categories

Figure 5 shows the distribution of *order by*, *group by*, and *having* statements in the generated queries. The mechanical queries are used here for comparison. Specific subsets thereof are also used for biased few-shot samples in the case of the 3-shot LLM prompts. The figure shows that (1) our approach can be as easily biased as the mechanical approach and (2) even though it was not considered initially, also generates *having* statements.

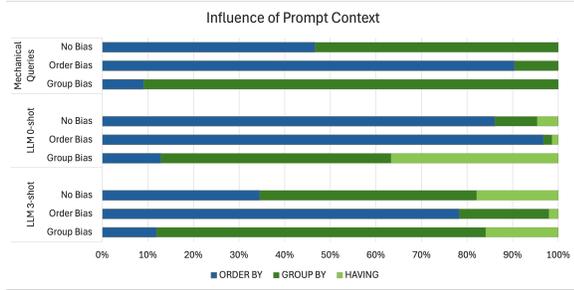

Figure 5: Query Distribution by Generation and Bias

## A.2 Query Execution Time & Cardinality

Figure 6 shows the execution times of the queries generated, as well as their cardinalities. Note that for queries generating no result, only those with a minimum runtime of 10s have been left in the generated data sets.

## A.3 Prediction of Query Execution Times

Table 2 shows the accuracy of the predicted execution times for each of the engines considered and their respective provisionings.

|  | $q_{mean}^{PrestoW1}$ | $q_{mean}^{PrestoW4}$ | $q_{mean}^{SparkW1}$ | $q_{mean}^{SparkW4}$ |
|---|---|---|---|---|
| $\text{Err}_{GNN+Mech}$ | 1.40 | 1.42 | 1.46 | 1.38 |
| $\text{Err}_{GNN+SDG}$ | 1.35 | 1.37 | 1.32 | 1.31 |

Table 2: LCM accuracy in predicting query execution times, extended for PrestoDB and Spark-SQL engines, with 1 and 4 workers, respectively.

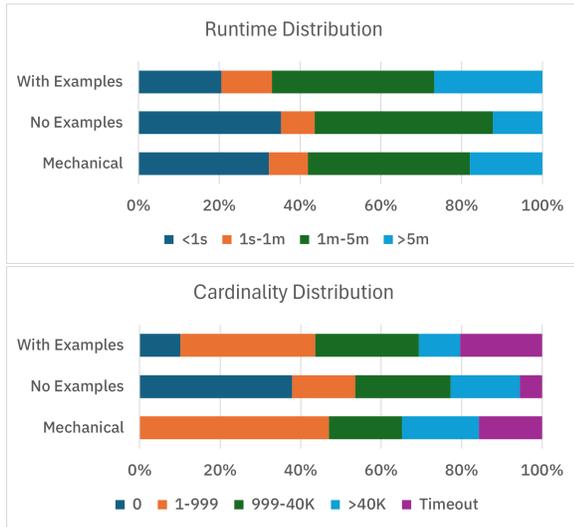

Figure 6: Runtime Characteristics of Generated Queries